\documentclass{article}

\usepackage{arxiv}

\usepackage{mathptmx}
\usepackage{amsmath}
\usepackage{amssymb}
\usepackage{amsthm}

\usepackage[utf8]{inputenc} 
\usepackage[T1]{fontenc}    
\usepackage{hyperref}       
\usepackage{url}            
\usepackage{booktabs}       
\usepackage{amsfonts}       
\usepackage{nicefrac}       
\usepackage{microtype}      
\usepackage{lipsum}

\usepackage{optidef}
\usepackage{algorithm}
\usepackage{algpseudocode}
\usepackage{algorithmicx}
\usepackage{enumitem}
\usepackage{tikz} 

\usepackage{caption}
\usepackage{subcaption}
\usepackage{centernot}
\usepackage{xcolor}

\newtheorem{Theorem}{Theorem}
\newtheorem{Lemma}{Lemma}
\newtheorem{Proposition}{Proposition}
\newtheorem{Definition}{Definition}
\newtheorem{Admissibility}{\text{Admissibility Rule}}
\newtheorem{Remark}{Remark}

\newcommand\independent{\protect\mathpalette{\protect\independenT}{\perp}}
\def\independenT#1#2{\mathrel{\rlap{$#1#2$}\mkern2mu{#1#2}}}
\newcommand{\notindependent}{\centernot{\independent}}

\DeclareMathAlphabet{\mathcal}{OMS}{cmsy}{m}{n}
\newcommand{\argmin}{\mathop{\mathrm{argmin}}}

\title{A novel framework extending cause-effect inference methods to multivariate causal discovery}

\author{
 Hongyi Chen \\
  Tilburg University \\
  the Netherlands \\
   \And
 Maurits Kaptein\\
  Tilburg University\\
  the Netherlands \\
}

\begin{document}
\maketitle
\begin{abstract}
We focus on the extension of bivariate causal learning methods into multivariate problem settings in a systematic manner via a novel framework. It is purposive to augment the scale to which bivariate causal discovery approaches can be applied since contrast to traditional causal discovery methods, bivariate methods render estimation in the form of a causal Directed Acyclic Graph (DAG) instead of its complete partial directed acyclic graphs (CPDAGs). To tackle the problem, an auxiliary framework is proposed in this work so that together with any bivariate causal inference method, one could identify and estimate causal structure over variables more than two from observational data. In particular, we propose a local graphical structure in causal graph that is identifiable by a given bivariate method, which could be iteratively exploited to discover the whole causal structure under certain assumptions. We show both theoretically and experimentally that the proposed framework can achieve sound results in causal learning problems.
\end{abstract}

\section{Introduction}\label{sec:intro}
Inferring causal relationships between variables from observational data has drawn great research interests, particularly in the fields of deep learning, medicine, and marketing, where the explanatory reasoning of predictive models has gained increasing attention. since in fields like deep learning, medicine, marketing etc, the explanatory reasoning of predictive models has come under increasing attention. The need to infer from a causal model is paramount for answering interventional or counterfactual questions, as inference from a non-causal model can lead to erroneous results. This will enable estimated causal relations to resemble the underlying true causal mechanism as closely as possible. Over the last few decades, numerous ideas and works have been discussed and published in this area, with the paper by \cite{glymour2019review} providing a comprehensive overview.

The school of thought built upon probabilistic graphical models, also known as structural equation models \cite{pearl2009causality,spirtes2000causation}, has produced a range of algorithms and methods, derived from different statistical methodologies, for causal structure learning in interdisciplinary research. One group of causal discovery algorithms that employ the Markov property encoded by conditional independence are theoretically sound. However, they are limited to identifying only the Markov equivalence class of the true causal graph instead the true causal graph itself unless additional assumptions such as non-linearity or additive noises are imposed. Another group of methods rely on the asymmetry induced by causal relations in the innate complexity discrepancy between variables. However, the majority of works in this group have focused on bivariate cases, whereas multivariate data, a more general and realistic setting, has not been adequately addressed. Given the research gap between these two groups of approaches, we aim to provide insight into how to generalize any bivariate methods into new ones that can be applied uniformly and systematically to multivariate data.

It may appear to be a simple task to extend bivariate methods to multivariate data, but in fact, the opposite is true. Some researchers have proposed that any edge in a complex network can be inferred by simply examining its two vertices independently \cite{goudet2018learning} while others advocate for conditioning on all remaining variables during inference \cite{lopez2015towards}. The main issue with the former approach is that its justification lies on the algorithmic Markov condition \cite{janzing2010causal}, which may no longer hold when the two endpoints are confounded by other variables in the network. On the other hand, although the latter approach is theoretically sound, it becomes impractical and unreliable when the number of conditioned variables increases, leading to multiple testing problems.

In light of the issues discussed earlier, we propose a new auxiliary framework for extending bivariate methods to multivariate causal inference. This framework allows for a systematic exploration of various causal discovery methods that focus on two variables, and can be expanded to include multivariate data under certain conditions. A key aspect of this framework is the use of a novel graphical criterion to determine the causal orders of all the variables. Specifically, we utilize a graphical criterion that compares the marginalized graph over an edge with the true causal graph to sequentially obtain the causal orders of each variable. We provide a theoretical analysis of our framework, including a general set of assumptions (admissibility rules) that ensure the correctness of individual bivariate methods, along with provable consistency results and considerations of computational complexity.

Indeed, our auxiliary framework provides myriad of benefits. Not only does it expand the range of applications for bivariate methods, but it also generates a family of new methods that can be useful in exploring a variety of inquiries, such as discovering confounders, causal relationships, and the orientation of edges in a partially directed graph.

The paper offers following important contributions:

\begin{itemize}
    \item  Firstly, it proposes a systematic approach for extending any bivariate causal inference method to the multivariate context, which is achieved through an auxiliary framework based on a newly proposed graphical criterion. 
    
    \item Secondly, the framework yields a new family of causal discovery algorithms that can be used in conjunction with traditional causal discovery approaches and bivariate methods. 
    
    \item Finally, the theoretical soundness of the framework has been demonstrated in cases where infinite data points are presented, and its computational feasibility has been demonstrated in practical applications.
\end{itemize}

\section{Background $\&$ Preliminaries}

We start with a brief introduction to the topics related to our research. In terms of notation, we denote random variables and random vectors with capital letters and bold upper-case letters, respectively, while their corresponding assigned values are represented by lowercase letters.

\subsection{Graph terminology}
A graph is an ordered pair defined as $\mathcal{G}=(V,E)$ consisting of a vertices set V and an edges set E. In the context of graphical modelling, vertices represent random variables and edges encode probabilistic and causal relations between vertices associated with them. For convenience, we use terms vertices and variables interchangeably.

A \textit{(un)directed} graph is a type of graph which contains only (un)directed edges. Otherwise, it is called a \textit{mixed} or \textit{partially directed} graph. In particular, a directed graph absent of \textit{directed cycles} is known as a \textit{Directed Acyclic Graph} (DAG) that could be transformed into an undirected graph by removing all edge directions, referred as \textit{skeleton} of the DAG. Two vertices are \textit{adjacent} if they are linked with an edge. A \textit{directed path} is a sequence of distinct vertices which are successively adjacent by edges of the same direction. If there is a directed path from vertex $X$ to $Y$, then $X$ is an ancestor of $Y$ while $Y$ is a descendant of $X$. Moreover, if such a directed path is an edge, we call $X$ a parent of $Y$ and $Y$ a child of $X$. The sets of parents, children, ancestors and descendants of a vertex $X$ in graph $\mathcal{G} $ are denoted as $\mathbf{PA}(\mathcal{G},X),\mathbf{CH}(\mathcal{G},X),\mathbf{AN}(\mathcal{G},X)$ and $\mathbf{DE}(\mathcal{G},X)$ accordingly.

A graphical criterion designated as \textit{d-seperation} \cite{lauritzen1996graphical, pearl2009causality} specifies conditional independence relationships of a DAG comprehensively. If joint distribution of $\mathbf{X}$, $\mathbb{P}(\mathbf{X})$, contains all the conditional independence relationships encoded by a DAG $\mathcal{G}$, the distribution is said to be Markovian to $\mathcal{G}$. On the other hand, a distribution $\mathbb{P}(\mathbf{X})$ is said to be faithful to a graph $\mathcal{G}$ if every conditional independence relation of $\mathbb{P}(\mathbf{X})$ is entailed by that in $\mathcal{G}$ \cite{spirtes2000causation}. If a distribution is both Markovian and faithful with respect to a DAG, we call the DAG a perfect map of the distribution.

A group of DAGs encoding the same $d$-separation statements forms a Markov equivalence class that can be uniquely represented by a Completed Partially Directed Acyclic Graphs (CPDAG) \cite{andersson1997characterization,chickering2002learning,pearl2009causal}. CPDAG contains directed edges that have the same orientation for every DAG in the equivalent class and undirected edges that have reversible orientations in the equivalent class. Every member within the same CPDAG would fit the data equally well and thus statistically equivalent, making themselves indistinguishable from causal discovery methodology embedded in independence testing or data fitting. 

\subsection{Structural Equation Models}

A structural equation model (SEM) determines the marginal distribution of each variable in the random vector $\mathbf{X}$=$(X_1, ..., X_n)$ corresponding to their DAG $\mathcal{G}$ by structural equations of the following form \cite{bollen2014structural}:
\[
X_i=f_i(\textbf{PA}(\mathcal{G},X_i),\epsilon_i)
\]
\noindent
where $\{\epsilon_i\}_{j=1,...,n}$ are mutually independent random noises and $\{f_i\}_{j=1,...,n}$ are real functions.

If a random vector ${X_1, ..., X_n}$ is generated according a SEM, we can factorize the density of the joint distribution as \cite{lauritzen1996graphical}:
\[
 p(x_1,...,x_n)=\prod_{1}^{n}p(x_i|\mathbf{PA}(\mathcal{G},x_i))
 \]
\noindent
It is clear that such a distribution is Markov to the DAG $\mathcal{G}$.

We now define an important concept called Pearl's \emph{do-intervention} \cite{pearl2009causality}. When we operate a do-intervention upon a variable $X_i$, we change the generating mechanism of $X_i$ and rewrite the SEM of $\mathbf{X}$ by updating the corresponding equation of $X_i$. This results in a new post-intervention distribution for 
$\mathbf{X}$. In particular, if the do-intervention fixes $X_i$ to a fixed point in the support of $X_i$, the joint density of $\mathbf{X}$ according to truncated factorization,
\[
 p(x_1,...,x_n|do(X_i=\hat{x_i}))=
  \begin{cases} 
   \prod \limits_{j\neq i} p(x_j|\mathbf{PA}(\mathcal{G},x_j)) & \text{if } X_i= x_i \\
   0      & \text{otherwise } 
  \end{cases}
\]

\subsection{Prior work on causal discovery}

Popular CPDAG learning methods include three categories: constraint-based methods, e.g. the PC algorithm \cite{spirtes2000causation}, utilizing conditional independence constraints embedded in the joint distribution of variables; score-based methods, e.g. Greedy equivalence search\cite{chickering2002learning}, employing a score function to search for optimal structures; and hybrid of the two, e.g. the Max-Min Hill-Climbing (MMHC) algorithm \cite{tsamardinos2006max}. By applying extra constraints on the SEM, the true DAG can be identified. Examples of this approach are Linear, Non-Gaussian, Acyclic causal Models (LiNGAM) \cite{shimizu2006linear,shimizu2009direct}; the additive noise model (ANM) \cite{hoyer2008nonlinear,mooij2009regression,peters2014causal}; and post-nonlinear model (PNL) \cite{zhang2009identifiability}.

On the other hand, a separate branch of inference methods built upon postulates such as the algorithmic Markov condition \cite{janzing2010causal,janzing2012information} in terms of \textit{Kolmogorov complexity} and \textit{independence of input and mechanism}. Even though these postulates are not verifiable yet, they offer philosophical building blocks, allowing us to interpret asymmetry between cause and effect on mathematical terms. Therefore, non-parametric algorithms following this principle, such as CURE \cite{sgouritsa2015inference}, IGCI \cite{janzing2012information}, CGNN \cite{goudet2018learning} and KCDC \cite{mitrovic2018causal} have been shown to be able to orient cause-effect pairs to various extents. Though some have sketched some intuitive ideas \cite{goudet2018learning,mitrovic2018causal,lopez2015towards} to apply these procedures to more than two variables, we believe it currently remains an open question on how to efficiently extend them to multivariable cases : it is exactly this question we aim to address in this paper.

\section{Extending bivariate methods to multivariate causal learning}
In this section, we begin with examining the idea of isolating a pair of variables for causal discovery in a multivariate setting. While this approach may seem intuitive, we find that it is not always reliable and is subject to specific conditions, as stated in Theorem 1. Theorem 1 is essential as it lays the foundation for enhancing the flawed approach with a reliable one presented in this study, for extending bivariate methods in multivariate data sets.

\subsection{Problem setting}
In this study, we examine a multivariate data scenario where a collection of $p$ random variables, denoted by $\textbf{X}:={X_1,...,X_p}$, are determined by a structural causal model (SCM) \cite{pearl2009causality}. This model characterises both their data-generating process and overall causal connections. Our work assumes that causal sufficiency and faithfulness hold. In other words, we assume that no unobserved confounding variables exist, and the conditional independencies in the distribution of $\textbf{X}$ align perfectly with d-separation relationships in the Directed Acyclic Graph (DAG), due to the satisfaction of both the Markov property and faithfulness.

Under the aforementioned settings, our focus is to develop an approach for inferring the complete or partial directed acyclic graph (DAG) of random variables $\textbf{X}$, using two key components: a dataset of $\textbf{X}$ and a bivariate method that can discern cause from effect between any two variables. Additionally, we are also interested in answering other related questions, such as identifying any further assumptions required for extending this approach; and determining the direction of undirected edges in the true CPDAG.

\subsection{Causality from Marginalisation}

Analogous to the way a graph is composed of many edges, the causal structure of a multivariate data set is constructed from the causal relations between each pair of variables, which can be detected by a bivariate method. However, it is generally incorrect to use a bivariate method on just one pair of variables to infer their causal relationship while disregarding the other variables. This is because such a procedure would effectively coincide with learning the causal direction with hidden, unmeasured variables that could be common causes of the pair, and this has been shown to potentially lead to incorrect causal conclusions, as in the well-known example of Simpson's Paradox. To establish the identifiability of a pair of variables using a bivariate method, we derive a sufficient condition that must be satisfied: the causal reconstruction based solely on the marginal distribution of the variables must correspond to the true causal relationship.

As suggested by authors in \cite{peters2014causal}, we provide a precise mathematical definition for the term "true causal graph" that we have been informally using, as outlined in Definition 1 according to \cite{peters2014causal}.

\begin{Definition}
\normalfont

Assume that random variables $\textbf{X}=\{X_1,...,X_p \}$ has $\mathcal{D}(\textbf{X})$ as its joint distribution as well as distributions $\mathcal{D}^{do(X_j:= \Tilde{N_j},j \in \textbf{J})}(\textbf{X})$ for all $\textbf{J} \subseteq \{1,...,\text{p}\} $ known, where $\Tilde{N_j}$ are random distributions. Then the DAG $\mathcal{G}$ is a true causal graph of $\mathbf{X}$ if conditions (i) and (ii) are satisfied.
\begin{enumerate}[label=(\roman*)]
    \item The joint distribution $\mathcal{D}(\textbf{X})$ is Markovian to $\mathcal{G}$
    \item for all $\textbf{J} \subseteq \{1,...,\text{p}\} $ and $\Tilde{N_j}, j\in \mathbf{J}$, the distribution $\mathcal{D}^{do(X_j:= \Tilde{N_j},j \in \textbf{J})}(\textbf{X})$ is identical to $\mathbb{P}^{\mathcal{G}; do(X_j):= \Tilde{N_j},j \in \textbf{J}}$
\end{enumerate}
\end{Definition}

Two graphical features pertaining to causality are introduced below.
\begin{Definition}
\normalfont
Given random variables $\textbf{X}=\{X_1,...,X_p \}$ whose joint distribution is generated by a SCM with DAG $\mathcal{G}$ as the causal graph, if $\mathbb{P}(X_j|do(X_i))=\mathbb{P}(X_j|X_i)$, then we have
\begin{enumerate}[label=(\roman*)]
    \item the DAG $X_i \rightarrow X_j, i,j \in \{1,...,\text{p}\}$ is said to be a \textbf{valid marginalisation} ;
    \item $X_i$ and $X_j$ are unconfounded and an edge between them is called an unconfounded edge if they are adjacent. 
\end{enumerate}

\end{Definition}

Lemma 1 details the relationship between the two previously defined features and how they pave the way for our seeking of a graphical criterion that connects the causal structure of two vertices' marginal distribution to that of the true causal graph. Theorem 1 demonstrates that being unconfounded is sufficient as such a criterion. All theoretical results are accompanied by proofs in the appendix.

\begin{Lemma}
\normalfont

For random variables $\textbf{X}=\{X_1,...,X_p \}$ with causal graph $\mathcal{G}$, then 
\begin{enumerate}[label=(\roman*)]
    \item any unconfounded edge of $\mathcal{G}$ is a valid marginalisation. 
    \item the DAG of two vertices: $X_i \rightarrow X_j, i,j \in \{1,...,p \}$ is the true causal graph of marginal distribution $\mathcal{D}(\{X_i,X_j\})$ if $X_i \rightarrow X_j, i,j$ is a valid marginalisation.
\end{enumerate}
\end{Lemma}

\begin{Theorem}
\normalfont
Suppose that an edge between $X_i$ and $X_j$, $i,j \in \{1,...,\text{p}\}$in causal graph $\mathcal{G}$ over random variables $\textbf{X}=\{X_1,...,X_p \}$ is unconfounded, then the causal direction of marginal distribution of  $X_i$ and $X_j$ coincides with the direction of $X_i$ and $X_j$ in the true causal graph $\mathcal{G}$.
\end{Theorem}

\subsection{Admissibility Rule}

As numerous causal discovery methods incorporate a priori assumptions concerning the structural equations' assignments, restraining the range of distribution families the variables can adopt, we endorse the notion and prescribe regulations to preserve the validity of utilizing bivariate methods on variables in multivariate data. These regulations, referred to as "Admissibility Rules," are presented below and elaborated on in Section 4. Only when these regulations are adhered to can causal approaches be considered reliable.

\begin{Admissibility}

\normalfont
For a non-confounding edge between $X_i,X_j,i,j\in \{1,...,\text{p}\}$ in \textbf{X}, it is only eligible to apply a bivariate method to $\mathcal{D}(X_i,X_j)$ if the marginal distribution fulfills functional premises required by the method.
\end{Admissibility}

Deviation from this admissibility condition can result in untenable inference. For example, if a non-Gaussianity method is applied to linear Gaussian variables without complying with the admissibility condition, it can yield false outcomes. In the following section, we investigate the compliance of Admissibility Rule 1 in the bivariate methods that we have reviewed in Section 2.3. Our findings demonstrate that there is a general consensus among these methods in adhering to the Admissibility Rule.

\begin{enumerate}
    \item For methods specifying particular choices of functional classes in the structural equations, such as LiNGAM, derivatives of ANM \cite{mooij2016distinguishing,shimizu2006linear,shimizu2009direct}, and PNL \cite{zhang2009identifiability}, it is easy to see LiNGAM is admissible. While for the rest, the marginal distribution of two variables generated by an ANM(PNL) cannot be expressed following a bivariate ANM(PNL) in general. A special case where it is possible to do so is when the varaibles are sampled according to  particular form of ANM called Causal Additive Models(CAM): $$X_j=\sum_{k\in \mathbf{PA}(X_j)}f_{jk}(X_k)+\epsilon_j$$ where $j \in \{1,...\text{p}\}$, $\epsilon_j$ is Gaussian and $f_{jk}$ is three times differential nonlinear functions.
    
    \item For the other group that are nonparametric without any assumptions on functional classes in the structural equations, they automatically satisfy Admissibility Rule 1.

\end{enumerate}

\section{Auxiliary Framework}

Building upon the theoretical findings established in Theorem 1, which establish the fundamental basis of the type of causal information that can be obtained through bivariate methods in multivariate scenarios, we introduce an\ textbf{auxiliary framework} for causal structure discovery, to be used in combination with any bivariate method.

In terms of structure, our proposed framework follows a top-down strategy in causal ordering \cite{de1986theories}, starting with the smallest causal ordering (root nodes) and working up to the largest (leaf nodes). The identification of root nodes is not only important but is also repeatedly applied inductive to infer further down the topological hierarchy. The framework is comprised of two main algorithms, the first of which determines the root nodes of the true causal graph, while the second enables comprehensive causal structure discovery. In most cases, both algorithms are able to unearth more causal information than CPDAG-oriented methods.

Furthermore, we conduct a theoretical analysis of the framework schemes to to demonstrate that they are both consistent and computationally feasible. For the sake of clarity, we assume the following conditions are met throughout this section unless otherwise indicated.

\begin{itemize}
    \item [A1] The true causal graph $\mathcal{G}$ is faithful to the data distribution $\mathcal{D}(\mathbf{X}) $.
    \item [A2] Causal sufficiency in $\mathbf{X}$
    \item[A3] (Conditional) Independence tests on the data sample are accurate.
    \item[A4] For bivariate data generated according to settings in Section 3.1, causal inference determined by the bivariate method is correct. 
\end{itemize}

A1 and A2 are commonly accepted assumptions in the field of causal discovery. In contrast, A3 and A4 are technical requirements that govern the population setting and the reliability of the statistical methods used. These conditions are crucial to establish the consistency of our framework.

\subsection{Root nodes determination}

Algorithm 1 outlines a procedure to identify the root nodes of the causal DAG $\mathcal{G}$ and their associated edge directions. The algorithm requires a sample of data $\textbf{X}$ and the skeleton of $\mathcal{G}$, as well as a given bivariate method. Initially, the algorithm transverse all undirected edges, estimating their causal direction through the bivariate method, and then constructs a directed graph $\mathcal{G}'$. Since not all root nodes of $\mathcal{G}'$ correspond to those in $\mathcal{G}$, the algorithm then applies independence tests and the bivariate method to identify any statistical dependence between any two root nodes in $\mathcal{G}'$, filtering out any false-positive root nodes. The remaining root nodes in $\mathcal{G}'$ comprise the root nodes of $\mathcal{G}$. Theorem 2 provides proof that this algorithm yields all and only the root nodes of $\mathcal{G}$.

\begin{algorithm}
\caption{}
\floatname{algorithm}{MegaAlgorithm}
\begin{algorithmic}[1]
\State \textbf{INPUT}: Skeleton $\mathcal{G}_s$ of a DAG of $n$ variables $\textbf{X}$, a bivariate method \textbf{BM} and a sample data set of $\textbf{X}$

    \State $S:=\{1,...,n\} R:=\{\emptyset\}$
    
    \For{k in $S$}
    
    \State apply \textbf{BM} to $X_k$ and each of its neighbours in $\mathcal{G}_s$ 
    
    \If{$X_k$ remains to be the parent of all directed edges inferred by \textbf{BM} }
    
    \For{j in $R$}
    
    \State Perform independence test between $X_k, X_j$
    
    \If{ $X_k \notindependent X_j$}
    
    \State apply \textbf{BM} to $X_k,X_j$
    
    \If{$X_j$ is the parent}
    
    \State \textbf{Break} from current loop
    
    \Else
    \State $R=R\backslash j$
    
    \EndIf
    \EndIf
    \EndFor
    \State $R=R\cup k$
    \EndIf
    \EndFor
\State \textbf{OUTPUT}: $X_i$ where $i \in R$ are root nodes and thus their edges' directions are oriented accordingly.

\end{algorithmic}
\end{algorithm}

In the following proposition, we provide a summary of theoretical results pertaining to the causal interpretation of using bivariate methods on two variables within $\mathbf{X}$.
\begin{Proposition}

\normalfont
Under A1 to A4 and Admissibility Rule 1,
\begin{enumerate}[label=\alph*.]
    \item A bivariate method might fail to accurately identify the causal direction of an edge if it is confounded.
    
    \item A bivariate method would be able to infer the correct causal direction of an edge if it is unconfounded.
\end{enumerate}

\end{Proposition}

\begin{Proposition}
\normalfont
Under A1 to A4 and Admissibility Rule 1, the causal direction of any edge connected to a root variable would be correctly inferred by a bivariate method. 
\end{Proposition}

An essential theorem regarding the accuracy of Algorithm 1 can be obtained once Propositions 1 and 2 have explained the objectives of utilising the given bivariate method.

\begin{Theorem}
\normalfont
Under A1 to A4 and Admissibility Rule 1, the output of Algorithm 1 contains all and only root nodes of the true causal graph over $\textbf{X}$.

\end{Theorem}

\begin{Remark}
\normalfont
Algorithm 1 not only facilitates the discovery of causal structures that are often obscured by Markov equivalence classes, but also embodies a concept that encompasses a mechanism for revealing causal relationships beyond root nodes, as illustrated in the latter part of our framework.
\end{Remark}

\subsection{Comprehensive Causal Discovery}

The second objective of our framework is to furnish a comprehensive algorithm that learns the entire causal graph using a bivariate method. Prior to delving into the specific steps outlined in Algorithm 2, we first introduce additional guidelines that the bivariate method must adhere to, as it is necessary for it to be applicable to conditional distributions. 

\begin{Admissibility}

\normalfont
For an adjacent edge $X_i,X_j,i,j\in \{1,...,\text{p}\}$ in \textbf{X} confounded by a disjoint variables set $\mathbf{Y}$ in $\mathbf{X}$, it is only eligible to apply a bivariate method to infer the direction of the conditional distribution $\mathcal{D}(X_i,X_j|\mathbf{Y})$, if the conditional distribution is identifiable by the method.

\end{Admissibility}

To ensure the correctness of our approach, we also need to verify whether bivariate methods satisfy Admissibility Rule 2 in general. We have observed that the methods discussed in Section 2.3, with the help of appropriate regression techniques, adhere to Admissibility Rule 2, as long as they follow Admissibility Rule 1, except for the IGCI method as discussed by \cite{janzing2010causal}. Therefore, most of the bivariate methods can be utilised for this purpose.

Algorithm 2 presents the pseudocode for a comprehensive causal graph discovery process in an iterative and inductive manner, following the same workflow as Algorithm 1. Unlike Algorithm 1 which terminates after root node retrieval, Algorithm 2 marks and designates root nodes as a batch of nodes with the same causal ordering before identifying the next batch of nodes in the subgraph. The marking process proceeds iteratively, excluding nodes that have already been marked until all nodes have been marked with a causal ordering. The different causal orderings for the batches defines the edge directions between nodes in different batches. It is worth noting that there are no edges between nodes within the same batch. As a result of this process, Algorithm 2 assigns a causal ordering to each node and orients all edges, thus specifying a DAG for the causal graph of $\mathbf{X}$.

Proposition 3 demonstrates the consistency of Algorithm 2, indicating that it will identify the true underlying DAG under regular assumptions.

\begin{Proposition}
\normalfont
Under A1 to A4 and Admissibility Rule 2, Algorithm 2 would infer the true causal DAG. 
\end{Proposition}

\begin{algorithm}[H]
   \caption{}
   \label{alg:example}
\begin{algorithmic}[1]
\State \textbf{INPUT}: Skeleton $\mathcal{G}_s$ of a DAG of $n$ variables $\textbf{X}$, bivariate method \textbf{BM} and a sample data set of $\textbf{X}$
   
   \State $S_0:=\{1,..,n\} S_1:= \{\emptyset \} R:= \{\emptyset \}$
   
   \For{$X_k$ where $k in S_0$ without any undirected edges}
   
   \State $S_0=S_0\backslash k$ $S_1=S_1 \cup k$
   \EndFor
   
   \Repeat
   \For{$k$ in $S_0$}
   \State apply \textbf{BM} to $X_k$ and each of its neighbours over undirected edges in $\mathcal{G}_s$, conditioned on a subset of $X_{S_1}$ which blocks all non-directed paths in $\mathcal{G}_s$ between $X_k$ and the particular neighbor. 
   
    \If{$X_k$ remains to be the parent of all directed edges inferred by \textbf{BM} }

    \For{j in $R$}
    
    \State Perform conditional independence test between $X_k, X_j$ given a subset of $X_{S_1}$, denoted as $S_b$, which blocks all non-directed paths between them in 
    $\mathcal{G}_s$ 
    
    \If{ $X_k \notindependent X_j | S_b$  }
    
    \State apply \textbf{BM} to $\mathcal{D}(X_j,X_k|\mathbf{S_b})$
    
    \If{$X_j$ is the parent}
    
    \State \textbf{Break} from current loop
    
    \Else
    \State $R=R\backslash j$
    
    \EndIf
    \EndIf
    \EndFor
    \State $R=R\cup k$
    \EndIf

    \EndFor
    \State Orient undirected edges linked to $X_i$, where $i \in R$ as directed edges pointing out of $X_i$ and update them to $\mathcal{G}_s$
    \State $S_0=S_0 \backslash R$ $S_1=S_1 \cup R$ $R=\{\emptyset \}$

   \Until $S_0=\{ \emptyset\}$
   
 \State \textbf{OUTPUT}: A fully directed graph,aka, a DAG
\end{algorithmic}
\end{algorithm}

\subsection{Computational Complexity}

To assess the computational cost of the framework, we have to consider the number of independence tests and the frequency with which the bivariate method is employed. The complexity of learning the CPDAG usually takes an upper bound of $\mathcal{O}(p^a)$, where a is constant equal to the size of the largest neighborhood. In the worst case scenario, Algorithm 2 performs $\mathcal{O}(p^3)$ iterations of bivariate method, while $\mathcal{O}(p^2)$ iterations of bivariate method are required if the graph is sparse. Though independence test and the bivariate method may scale with the sample size, the overall computational complexity of Algorithm 2 is polynomial in the number of variables $\text{p}$.

\section{Simulation Study}

Although the main contribution of this paper is a conceptual extension of a category of methods into a broader setup, it is important to demonstrate he practical application of the general framework using a particular bivariate method and, more crucially, to validate our approach with empirical studies. To this end, we assess the efficacy of a version of the auxiliary framework that has been implemented using synthetic data generated from a benchmark SCM as outlined in \cite{peters2014causal}.

In particular, we focus on triple-variable data sampled from a full DAG as ground truth on their causal relations. See in Figure 1a. As a standard scheme suggested in \cite{peters2014causal}, we generate $\mathbf{X}=\{X_1,X_2,X_3\}$ with a linear additive noise model with non-Gaussian error terms of the form:

    $X_1 =\epsilon_1$
    
    $X_2 = \beta_{21}X_1+\epsilon_2$
    
    $X_3 = \beta_{31}X_1+\beta_{32}X_2+\epsilon_3$

in which $\beta_{jk}$ are uniformly sampled from $[0.1,2]$ while  $\epsilon_i, i\in \{1,2,3\}$ are independent noise terms with distributions as  
$\psi_i \cdot \text{sgn}(N_i) \cdot |N_i|^{\alpha_i}$, where $N_i\stackrel{\text{iid}}{\sim} \mathcal{N}(0,1)$, $\psi_i \stackrel{\text{iid}}{\sim} \mathcal{U}([0.1,0.5])$ and $\alpha_i \stackrel{\text{iid}}{\sim} \mathcal{U}\{2,3,4\} $

\begin{figure}
    \centering
    \begin{subfigure}[b]{0.6\textwidth}
   \centering
    
        \begin{tikzpicture}[main/.style = {draw, circle}] 
           \node[main] (1) {$X_2$};
           \node[main] (2) [above right of=1] {$X_1$};
           \node[main] (3) [below right of=2] {$X_3$}; 
           \draw[->] (2) -- (1);
           \draw[->] (1) -- (3);
           \draw[->] (2) -- (3);

        \end{tikzpicture}     
    \caption{Underlying DAG representing causal structure of variables $X_1,X_2,X_3$ in the case study }
    \label{fig:my_label}

    \end{subfigure}
    \begin{subfigure}[b]{\textwidth}
    \centering
    \includegraphics[width=0.6\textwidth, height=8cm]{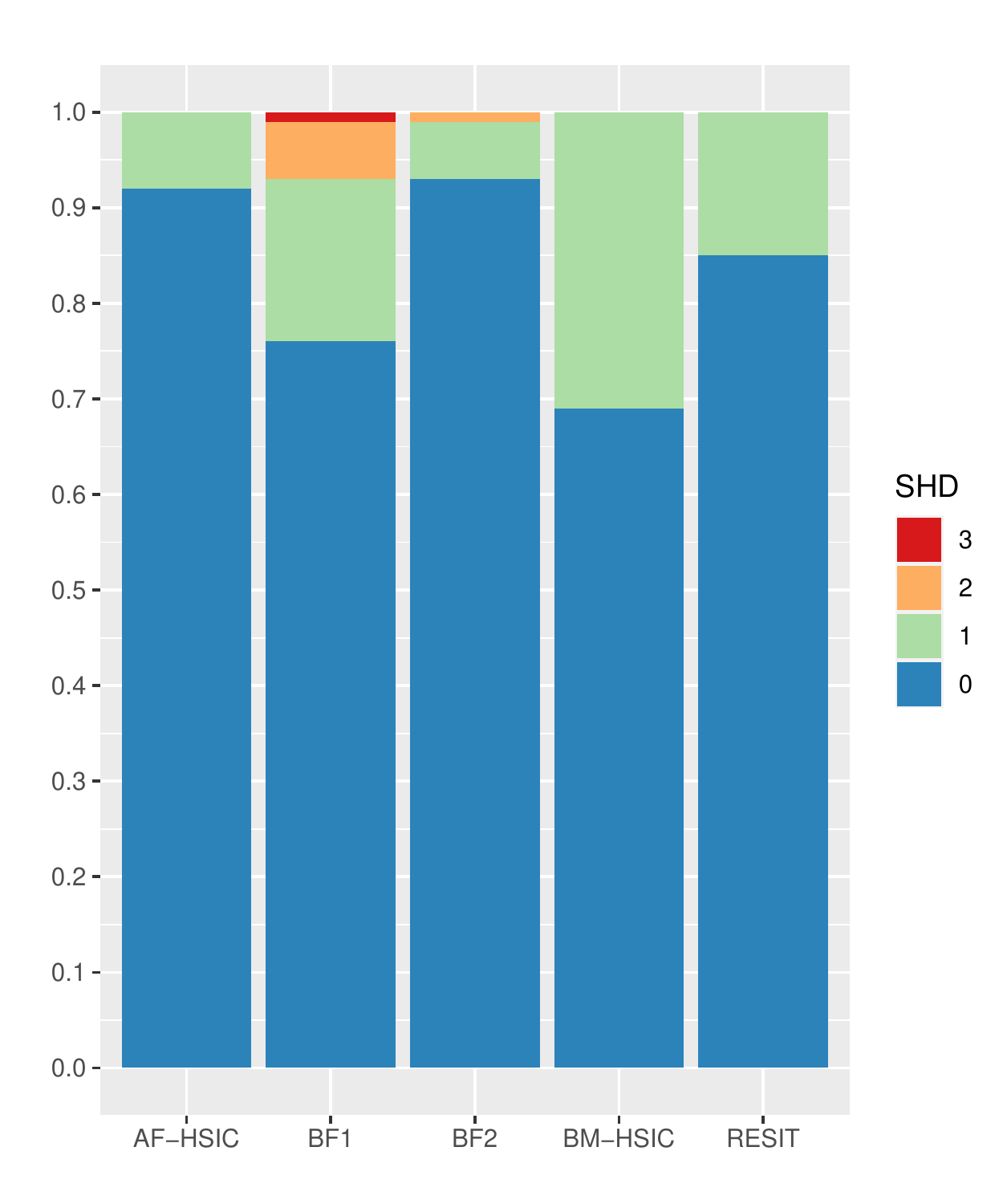}
    \caption{Bar plot of SHD between the estimated DAG and the ground truth DAG}
    \end{subfigure}
\caption{}
\end{figure}

We conduct 100 distinct experiments, each involving the random generation of parameter sets as depicted above. For each experiment, we prepare a sample of 100 tuples $(X_1,X_2,X_3)$. Our empirical analysis not only assesses the accuracy of our proposed framework in inferring causal structures from the observed data, but also compares its performance to that of existing state-of-the-art methods, including two brute-force based methods\cite{peters2014causal, pfister2018kernel}, regression followed by independence test (RESIT) \cite{peters2014causal}, and a pairwise ANM based method \cite{mooij2016distinguishing} with a procedure where the method is applied to edges independently, as suggested in \cite{goudet2018learning}.

In order to reduce ambiguity arising from the multitude of regression methods and independence measures that could potentially be used by all the methods in our study, we have opted to use the Hilbert Schmidt independence criterion (HSIC) \cite{gretton2005kernel} and linear regression as standard tools. Our intention is that this choice will increase the comprehensibility of the simulation results, enabling us to attribute the performance to the methods in comparison themselves rather than extraneous factors. Furthermore, all methods are provided with the causal graph skeleton.

\begin{Remark}
\normalfont

Brute-force methods can be quite effective when working with small graphs because they are enumerative methods, examining the entire set of possible directed acyclic graphs (DAGs) one by one. However, there is no consensus on how to evaluate all the candidate DAGs. In this study, we have selected two metrics, referred to as BF1 and BF2, to use for comparison, of which the former is suggested by \cite{peters2014causal} and the latter by ourselves, to select the best DAG as the output: $\hat{\mathcal{G}}= \argmin_{\mathcal{G}}\sum^p_{i=1}\Tilde{\mathcal{C}}(\text{res}_{i}^\mathcal{G},\text{res}_{-i}^\mathcal{G}) + \lambda |\text{edges}|$ or $ \hat{\mathcal{G}}= \argmin_{\mathcal{G}}\Tilde{\mathcal{C}}(\text{res}_{1}^\mathcal{G},...,\text{res}_{p}^\mathcal{G}) + \lambda |\text{edges}|$, where $\Tilde{\mathcal{C}}$ is a independence measure such as HSIC used in our paper, $\text{res}_{i}$ means residual of $X_i$ regressing on its parents, $\text{res}_{-i}$ denotes all such residuals except $X_i$. The regularization parameter $\lambda$ is often pruned accordingly but it is irrelevant to our work. As we can see from the empirical results later on, the choice of metric plays a significant role in the performance of brute-force methods. 
    
\end{Remark}

Algorithm 3 (See Appendix) gives a sketch on implementing the auxiliary framework with a HSIC-based bivariate method known as ANM-HSIC \cite{mooij2016distinguishing} where the independence measure, same as the score function $\hat{\mathcal{C}}$, is defined as the value of the empirical HSIC estimator itself between two data samples $\mathbf{u},\mathbf{v}$:
$\hat{\mathcal{C}}(\mathbf{u,v})=\widehat{\text{HSIC}}_{\phi_{l(\mathbf{u})},\phi_{l(\mathbf{v})}}(\mathbf{u,v})$
where $\phi_l$ is a kernel with parameters $\textit{l}$, that are estimated from the data. We name this instantiation AF-HSIC.

\begin{Remark}
\normalfont

One important point to consider is that while the auxiliary framework can extend the bivariate methods to multivariate settings, it does not enhance the accuracy of the chosen bivariate method and that mistakes made by the bivariate method may be propagated during the iterations of the auxiliary framework, much like the independence testing of the PC algorithm\cite{spirtes2000causation}. When dealing with dense graphs, it is recommended to take precautions such as verifying the acyclicity of the inferred DAG. Nonetheless, our simulated experiments in sparse graph scenarios did not encounter such robustness issues.
\end{Remark}

\begin{table}[h]
    \centering
    \begin{tabular}{p{0.15\textwidth} p{0.15\textwidth} p{0.15\textwidth} p{0.15\textwidth} p{0.15\textwidth}}
      \hline
      AF-HSIC & BF1 & BF2 & RESIT & BM-HSIC\\
      \hline\hline
      $\mathbf{0.08}$ $\pm$ $\mathbf{0.27}$ & 0.32 $\pm$ 0.63 & 0.08 $\pm$ 0.31 & 
      0.15 $\pm$ 0.36 & 0.31 $\pm$ 0.46\\
      \hline
    \end{tabular}

    \caption{Mean and standard deviation of SHD between estimated DAG and the true one over 100 simulated experiments}
    \label{tab:my_label}
\end{table}

Table 1 and the bar plot in Figure 1(b) present a summary of our experiment results. We assessed the accuracy of identifying the true DAG and the Structural Hamming distance (SHD) \cite{tsamardinos2006max} of the estimated DAG from various methods to the true causal graph. The SHD is defined as the number of edges in disagreement between the true DAG and the estimated DAG. Our proposed method (AF-HSIC) and the Brute-force method, using a measure we proposed, achieved the highest accuracy of 92$\%$, with only one incorrect edge in other cases. In contrast, applying the pairwise method to edges separately produced particularly poor results in the simulation, presumably due to confounding effects in the true DAG. Overall, our AF-HSIC outperforms all other methods, including the brute-force method, in linear non-Gaussian settings, while remaining computationally feasible.

Additionally, we have employed our framework with KCDC\cite{mitrovic2018causal} to simulation settings of $k$ variables, where $k=\{5,10\}$, under more complicated noises such as multiplicative or periodic. The results again have shown soundness and stellar performance of our framework. Details are given in the supplementary material. 
\section{Real-world data}
To determine the performance on real-world data sets, we apply an instantiation algorithm of our framework, AF-HSIC accompanied by smoothing splines as regression methods, to the protein network problem \cite{sachs2005causal}, labelled as Anti-CD3/CD28 dataset with 853 observational data points. A consensus network in the form of a DAG is provided by \cite{sachs2005causal}, which is regarded as the true causal structure of the dataset for comparison in this paper. Given the same background knowledge which is the underlying skeleton of the causal graph, our method identifies 16 out of 18 edges accurately using the data, thus performing better than other methods such as GES, CAM and CGNN according to results reported in \cite{goudet2018learning}. Table 2 shows the details, part of which is excerpt from \cite{goudet2018learning}.

\begin{table}[h]
    \centering
    \begin{tabular}{p{0.2\textwidth} p{0.2\textwidth} p{0.2\textwidth} p{0.2\textwidth} }
      \hline
      AF-HSIC & GES & CAM & CGNN \\
      \hline\hline
      $\mathbf{2}$ & 12.1&8.5&4.3\\
      \hline
    \end{tabular}

    \caption{Results of different methods for the orientation of the protein network given true skeleton in terms of SHD with the true causal graph.}
    \label{tab:my_label}
\end{table}
\section{Conclusion}

In this paper, we have introduced an auxiliary framework aimed at extending bivariate causal discovery methods to multivariate cases, resulting in a novel family of causal inference methods that unify both constraint-based methods and cause-effect inferences. Furthermore, we have demonstrated the framework's desirable theoretical properties in terms of consistency and computational costs under regular assumptions. We have also provided an instantiation of this framework over triple variables and evaluated it on diverse synthetic datasets against several state-of-the-art causal discovery methods, as well as on real-world data, where our algorithm outperformed other methods. The experiments presented in this paper not only support the theoretical analysis but also highlight the auxiliary framework's potential as a powerful tool for observational data aimed at causal discovery.

\newpage
\appendix

\section{Proofs}

\subsection{Proof of Lemma 1 }
\textit{Proof.} (i) Wlog, let $X_i \rightarrow X_j$ be an unconfounded edge in $\mathcal{G}$. By Definition 2(ii), we have $\mathbb{P}(X_j|do(X_i))=\mathbb{P}(X_j|X_i)$, which directly shows $X_i \rightarrow X_j$ is a valid marginalization from Definition 2(i).

(ii)To demonstrate $X_i \rightarrow X_j$ is the true causal graph, we need to show conditions in Definition 1 are satisfied. Part (i) of Definition 1 is trivial. To verify Definition 1(ii) holds, we need 
\begin{equation}
    \mathcal{D}^{do(X_j:= \Tilde{N_j},j \in \textbf{J})}(\textbf{X})=\mathbb{P}^{\mathcal{G}; do(X_j):= \Tilde{N_j},j \in \textbf{J}}
\end{equation}
where \textbf{J} is any subset of $\{i,j\}$.

Clearly, for subsets $ \mathbf{J}= \emptyset \text{ or } \{i,j\}$, we observe that both sides of (1) equal to the marginal distribution of $\{X_i,X_j\}$ that has not been intervened or totally intervened respectively. 

When $ \mathbf{J}=\{j\}$, the intervention only occurs on the child node independently of its parent node. Therefore, both sides of (1) can be written in the factorized product of two independent variables with known distributions, which shows the equality. 

For $ \mathbf{J}=\{i\}$, as $X_i \rightarrow X_j$ is a valid marginalization, RHS of (1) = $\mathbb{P}(X_j|do(X_i))*\mathbb{P}(do(X_i))$
using Definition 2(i).

While based on truncated factorization \cite{pearl1994belief}, LHS of (1)=  $\mathbb{P}(X_j|do(X_i))*\mathbb{P}(do(X_i))$. Therefore, (1) holds when $ \mathbf{J}=\{i\}$.

By enumeration of all subsets of $\{i,j\}$, we have shown (1) stands. Therefore, $X_i \rightarrow X_j$ satisfies Definition 1 so that it is the true causal graph over $\mathcal{D}(\{X_i,X_j\})$. \qedsymbol

\subsection{Proof of Theorem 1}
\textit{Proof.} Suppose that wlog the edge between $X_i,X_j$ in $\mathcal{G}$ is $X_i \rightarrow X_j$. From Lemma 1(i) and 1(ii), $X_i \rightarrow X_j $ is a valid marginalization and thus is the true causal graph over $\mathcal{D}(\{X_i,X_j\})$. Hence, two causal directions are the same. \qedsymbol

\subsection{Proof of Proposition 1}
\textit{Proof.} (a) Counterexamples can be inferred from the simulation study in Section 5.

(b)From Theorem 1, we know that the causal direction between two unconfounded nodes $X_i,X_j$ from their marginal distribution is the same as the edge orientation in the true causal graph $\mathcal{G}$. Consequently, to infer the direction between $X_i,X_j$ in $\mathcal{G}$ is equivalent to infer that from the marginal distribution of $X_i,X_j$, which can accurately obtained from applying the \textbf{BM} on them. Therefore, the output of the procedure described above produces the correct direction between $X_i,X_j$ in $\mathcal{G}$.  \qedsymbol

\subsection{Proof of Proposition 2}

\textit{Proof.} Wlog, let $X_i$ be any root node in $\mathcal{G}$. We will first show that any edge originates from $X_i$ is unconfounded. 

Suppose we have an edge $X_i\rightarrow X_j$ originating from a root node. Since an empty set $\emptyset$ doesn't contain any descendants of $X_i$ nor is there any backdoor paths into $X_i$ in $\mathcal{G}$, we have $\emptyset$ as a valid adjustment set for $(X_i,X_j)$ by Pearl's backdoor criterion \cite{pearl1994belief}:
$$\mathbb{P}(X_j|do(X_i))=\mathbb{P}(X_j|X_i)$$
Hence $X_i\rightarrow X_j$ is unconfounded by definition 2. Followed by Proposition 1(b) which states any unconfounded edge can be correctly discovered by the \textbf{BM}, proposition 2 is proved. \qedsymbol

\subsection{Proof of Theorem 2}

\textit{Proof.} We observe that due to Proposition 2, all root nodes could be identified via Algorithm 1, which means there's no root nodes missing in the finding of Algorithm 1. 

As for the proof of exclusiveness, suppose a node $X_n$ in the output of Algorithm 1 is not a root node, then there is a root node, denoted as $X_r$ such that  $X_n \notindependent X_r$, which indicates that $X_n$ is a descendent of $X_r$, i.e. there is a directed path from $X_r$ towards $X_n$. Hence, $X_r \rightarrow X_n$ is a valid marginalization since $X_r$ is a root node. If we apply a bivariate method to $\{X_n,X_r\}$, the correct direction would be inferred by Proposition 2, which is exactly a step described in Algorithm 1. Therefore, we have reached a contradiction of having $X_n$ in the output whereas $X_n$ has been identified as a descendent of another node, a forbidden case prescribed by Algorithm 1.
\qedsymbol 

\subsection{Proof of Proposition 3}

\textit{Proof.} Under assumptions stated in the Proposition premise, Algorithm 2 iteratively tries to identify all the 'root nodes' on the causal hierarchy ranking excluding the uncovered nodes from previous iterations. Since on each iteration, Algorithm 2 employs Algorithm 1 that could correctly identify all root nodes excluding the already discovered, Algorithm 2 could accurately identify 'root nodes' in each loop on the remaining variables. Hence, after all nodes have been identified, we have the true full causal graph inferred. \qedsymbol


\section{Algorithm for simulation of three-variable causal inference}
\begin{algorithm}[H]
   \caption{}
   \label{}
\begin{algorithmic}

    \State \textbf{INPUT}: Data sampled from $\{X_1,X_2,X_3\}$ tuple;  Score function $\hat{\mathcal{C}}$ based on HSIC, measuring independence between two samples.

   \For{$i,j \in \{1,2,3\}, i>j$} 
   \State Compute residuals of regressing $X_i$ on $X_j$ $\hat{\mathbf{e}}(X_i \sim X_j)$ as well as regressing $X_j$ on $X_i$ $\hat{\mathbf{e}}(X_j \sim X_i)$.

   \If{
   \begin{align*}
       \hat{\mathcal{C}}(X_i,\hat{\mathbf{e}}(X_j \sim X_i)) &< \hat{\mathcal{C}}(X_j,\hat{\mathbf{e}}(X_i \sim X_j))\\
       \hat{\mathcal{C}}(X_i,\hat{\mathbf{e}}(X_k \sim X_i)) &< \hat{\mathcal{C}}(X_k,\hat{\mathbf{e}}(X_i \sim X_k))\\      
   \end{align*}}
   \State Orient edges between $X_i,X_j$ and $X_i,X_k$ as $X_i \rightarrow X_j, X_i \rightarrow X_k$ respectively.
   \State Compute residuals of regressing $X_k$ on $X_j,X_i$ $\hat{\mathbf{e}}(X_k \sim X_j + X_i)$ as well as regressing $X_j$ on $X_k,X_i$ $\hat{\mathbf{e}}(X_j \sim X_k + X_i)$. 
   
   \If{
   $\hat{\mathcal{C}}(\hat{\mathbf{e}}(X_k \sim X_j + X_i),\hat{\mathbf{e}}(X_j \sim X_i)) < \hat{\mathcal{C}}(\hat{\mathbf{e}}(X_j \sim X_k + X_i),\hat{\mathbf{e}}(X_k \sim X_i))$}
   
   Orient edge between $X_j,X_k$ as $X_k \rightarrow X_j$
   
   \Else
   
   \State Orient edge between $X_j,X_k$ as $X_j \rightarrow X_k$
   
   \EndIf
   \EndIf
   
   \EndFor
   
   \State \textbf{OUTPUT} Estimated DAG for $\{X_1,X_2,X_3\}$

\end{algorithmic}
\end{algorithm}

\section{Additional simulation results on multivariate data}

We randomly generate 100 DAGs of $k=\{5,10\}$ vertices with maximum degree of 3 respectively, some of which could be identical. For each DAG generated, we simulate data set of 100 points according to the following data generating mechanism:
$$X_i=\sum_i^{n_i} f_{m_j}(\textbf{Pa}(X_i)_j)+\epsilon_i$$
, where $m_j=\{1,2,3\}$, $j \in\{1...n_i\}$, $n_i$ is the degree of $X_i$, $\epsilon_i$ are iid $\mathcal{N}(0,1)$ and $f_1(x)=x^3+x, f_2(x)=log(x+10)+x^6, f_3(x)=\sin(10x)+\exp(3x)$. Hyperparameter $m_j$ is randomly chosen for each $j$.

From the true CPDAG, we use \textbf{AF-KCDC}, in which we apply our algorithm to KCDC\cite{mitrovic2018causal}, to obtain estimated DAG from synthetic data sets. \textbf{AF-KCDC} achieves 100$\%$ estimation accuracy in depicted empirical studies over 100 DAGs for both $k=\{5,10\}$.


\bibliographystyle{unsrt}  
\bibliography{template}

\end{document}